\documentclass [final,1p,times]{elsarticle} 
\usepackage{graphicx} 
\usepackage{amssymb} 
\usepackage{amsthm} 
\usepackage{lineno} 

\journal{Nuclear Physics A} 
\begin{document} 

\begin{frontmatter} 


\title{Quark Phase Transition Parameters and
$\delta$-Meson Field in RMF Theory}

\author{G.B.Alaverdyan}

\address{Yerevan State
University, A.Manoogyan str.1, Yerevan 0025, Armenia}

\begin{abstract} 
The deconfinement phase transition from hadronic matter to quark
matter in the interior of compact stars is investigated. The
hadronic phase is described in the framework of relativistic
mean-field (RMF) theory, when also the scalar-isovector
$\delta$-meson effective field is taken into account. To describe
a quark phase the MIT bag model is used. The changes of the mixed
phase threshold parameters caused by the presence of
$\delta$-meson field are investigated.
\end{abstract} 
\end{frontmatter} 
\linenumbers 
\section{Deconfinement phase transition parameters}\label{}
The modern concept of hadron-quark phase transition is based on
the feature of that transition, that is the presence of two
conserved quantities in this transition: baryon number and
electric charge\cite{Gl92}. It is known, that depending on the
value of surface tension, $\sigma_{s}$, the phase transition of
nuclear matter into quark matter can occur in two scenarios
\cite{Heis93}: ordinary first order phase transition with a
density jump (Maxwell construction), or formation of a mixed
hadron-quark matter with a continuous variation of pressure and
density \cite{Gl92}. Uncertainty of the surface tension values
does not allow to determine the phase transition scenario, taking
place in realty. In our recent paper \cite {Al09a} in the
assumption that the transition to quark matter is a usual
first-order phase transition, described by Maxwell construction,
we have shown that the presence of the $\delta$-meson field leads
to the decrease of transition pressure $P_{0}$, of baryon number
densities $n_{N}$ and $n_{Q}$. In this article we investigate the
deconfinement phase transition, when the transition proceeds
through a mixed phase.

For description of hadronic phase we use the relativistic
Lagrangian  density of many-particle system consisting of
nucleons, $p$, $n$, and exchanged mesons
$\sigma,~\omega,~\rho,~\delta$:
\begin{eqnarray}{\cal L}_{\sigma\omega\rho\delta}
(\sigma(x),\omega _{_{\mu }}(x),\vec{\rho }_{_{\mu }}(x),
\vec{\delta}(x)) ={ \cal L}_{\sigma\omega \rho}(\sigma(x),\omega
_{_{\mu }}(x),\vec{\rho }_{_{\mu }}(x))-U(\sigma(x))+ {\cal
L}_{\delta}(\vec{\delta}(x)),
\end{eqnarray}where $\cal L_{\sigma\omega \rho}$ is the linear part of
relativistic Lagrangian density without $\delta$-meson field
\cite{Gl00}, $U(\sigma)=\frac{b}{3}m_{N}\left( g_{\sigma }\sigma
\right) ^{3}+\frac{c}{4}\left( g_{\sigma }\sigma \right) ^{4}$ and
$\cal L_{\delta}$$(\vec{\delta})=g_{\delta } \bar {\psi}_{N}
\vec{\tau }_{N} \vec{\delta }\psi_ {N}+\frac{1}{2}\left(\partial
_{\mu }\vec{\delta}\partial ^{\mu}
\vec{\delta}-m_{\delta}\vec{\delta}^{2}\right)$ are the
$\sigma$-meson self-interaction term and contribution of the
$\delta$-meson field, respectively. This Lagrangian density (1)
contains the meson-nucleon coupling constants, $g_{\sigma },~
g_{\omega },~g_{\rho },~g_{\delta}$ and also parameters of
$\sigma$-field self-interacting terms, $b$ and $c$. In our
calculations we take
$a_{\delta}=\left(g_{\delta}/m_{\delta}\right)^2=2.5$ fm$^2$ for
the $\delta$ coupling constant, as in \cite{LG_GC}. Also we use
$m_{N}=938.93$ MeV for the bare nucleon mass,
$m_{N}^{\ast}=0.78~m_{N}$ for the nucleon effective mass,
$n_{0}=0.153$ fm$^{-3}$ for the baryon number density at
saturation, $f_{0}=-16.3$ MeV for the binding energy per baryon,
$K=300$ MeV for the incompressibility modulus, and
$E_{sym}^{(0)}=32.5$ MeV for the asymmetry energy. Five other
constants, $a_{i}=\left(g_{i}/m_{i}\right)^2$ $(i=\sigma,~
\omega,~ \rho)$, $b$ and $c$, then can be  numerically determined:
$a_{\sigma}=\left(g_{\sigma}/m_{\sigma}\right)^2=9.154$ fm$^2$,
$a_{\omega}=\left(g_{\omega}/m_{\omega}\right)^2=4.828$ fm$^2$,
$a_{\rho}=\left(g_{\rho}/m_{\rho}\right)^2=13.621$ fm$^2$,
$b=1.654\cdot10^{-2}$ fm$^{-1}$, $c=1.319\cdot10^{-2}$. If we
neglect the $\delta$ channel, then $a_{\delta}=0$ and
$a_{\rho}=4.794$ fm$^{2}$. The knowledge of the model parameters
makes it possible to solve the set of four equations and to
determine the re-denoted mean-fields, $\sigma \equiv
g_{\sigma}\bar {\sigma}$, $\omega \equiv g_{\omega}\bar
{\omega_{0}}$, $\delta \equiv g_{\delta}\bar{\delta}^{\small(3)}$,
and $\rho \equiv g_{\rho}\bar {\rho_{0}}^{(3)}$, depending on
baryon number density $n$ and asymmetry parameter
$\alpha=(n_n-n_p)/n$. The standard QHD procedure allows to obtain
expressions for energy density $\varepsilon(n,\alpha)$ and
pressure $P(n,\alpha)$. The results of our analysis show that the
scalar - isovector $\delta$-meson field inclusion increases the
value of the energy per nucleon. This change is strengthened with
the increase of the nuclear matter asymmetry parameter, $\alpha$.
The $\delta$-field inclusion leads to the increase of the EOS
stiffness of nuclear matter due to the splitting of proton and
neutron effective masses, and also due to the increase of
asymmetry energy (for details see Ref.\cite{Al09b}).

To describe  the quark phase the MIT bag model is used, in which
the interactions between $u,~d,~s$  quarks inside the bag are
taken in a one-gluon exchange approximation \cite{Far84}. We
choose $m_{u} = 5$ MeV, $m_{d} = 7$ MeV and $m_{s} = 150$ MeV for
quark masses,  $B=60$ MeV/fm$^3$ for bag parameter and
$\alpha_{s}=0.5$ for the strong interaction constant.
\begin{table}[ht]
  \centering
  \caption{The Mixed phase parameters with ($\sigma\omega\rho\delta$) and without ($\sigma\omega\rho$) $\delta$-meson field.}
  \label{T1}
\begin{tabular}
[c]{|c|c|c|c|c|c|c|}\hline  & $n_{N}$ & $n_{Q}$  & $P_{N}$ &
$P_{Q}$ & $\varepsilon_{N}$ & $\varepsilon_{Q}$ \\
 & fm$^{-3}$ & fm$^{-3}$ & MeV/fm$^{3}$ & MeV/fm$^{3}$ &
MeV/fm$^{3}$ & MeV/fm$^{3}$
\\\hline $\sigma\omega\rho$ & 0.072 & 1.083 & 67.728 & 1280.889 &
0.336 & 327.747 \\\hline $\sigma\omega\rho\delta$ & 0.077 & 1.083
& 72.793 & 1280.884 & 0.434 & 327.745\\\hline
\end{tabular}
\end{table}

Table 1 represents the parameter sets of the mixed phase both with
and without $\delta$-meson field. It is shown that the presence of
$\delta$-field alters threshold characteristics of the mixed
phase. The lower threshold parameters, $n_{N}$, $\varepsilon_{N}$,
$P_{N}$, are increased, meanwhile the upper ones, $n_{Q}$,
$\varepsilon_{Q}$, $P_{Q}$, are slowly decreased. For EOS, used in
this study, the central pressure of the maximum mass neutron stars
is less than the mixed phase upper threshold $P_Q$. Thus, the
corresponding hybrid stars do not contain pure strange quark
matter core.

\end{document}